\documentclass[twocolumn,aps,prl,superscriptaddress,nofootinbib]{revtex4-2}

\usepackage{graphicx}
\graphicspath{{figures/}{fig/}}
\usepackage{amsmath}
\usepackage{bm}
\usepackage{slashed}
\usepackage{amsfonts}
\usepackage{color}
\usepackage{extarrows}
\usepackage{multirow}
\usepackage{hyperref}

\newcommand{\ben}{\begin{eqnarray}}
\newcommand{\een}{\end{eqnarray}}

\newcommand{\bef}{\begin{figure}[!htp]}
\newcommand{\eef}{\end{figure}}

\newcommand{\bea}{\begin{eqnarray}}
\newcommand{\eea}{\end{eqnarray}}

\def\ba{\begin{linenomath*}\begin{equation}}
\def\ea{\end{equation}\end{linenomath*}}

\newcommand{\sect}[1]{{\it \textbf{#1.} --- }}
\newcommand{\mi}{\mathrm{i}}
\newcommand{\md}{\mathrm{d}}

\allowdisplaybreaks

\newcommand{\state}[4]{{^{#1}\hspace{-0.6mm}#2_{#3}^{[#4]}}}

\newcommand\CSaSz{\state{1}{S}{0}{1}}

\newcommand\CScSa{\state{3}{S}{1}{1}}
\newcommand\CScPz{\state{3}{P}{0}{1}}

\newcommand\CScPj{\state{3}{P}{J}{1}}
\newcommand\COaSz{\state{1}{S}{0}{8}}

\newcommand\COcSa{\state{3}{S}{1}{8}}
\newcommand\COcPz{\state{3}{P}{0}{8}}
\newcommand\COcPa{\state{3}{P}{1}{8}}
\newcommand\COcPb{\state{3}{P}{2}{8}}
\newcommand\COcPj{\state{3}{P}{J}{8}}

\begin{document}

\title{\texorpdfstring{\boldmath Resolving the universality and polarization puzzles in $J/\psi$ hadroproduction}{Resolving the universality and polarization puzzles in J/psi hadroproduction}}


\author{An-Ping Chen}
\email{chenanping@jxnu.edu.cn}
\affiliation{ School of Physics, Jiangxi Normal University, Nanchang 330022, China}
\affiliation{Jiangxi Provincial Key Laboratory of Advanced Electronic Materials and Devices,
Nanchang 330022, China}
\author{Jun Guo}
\email{jguo_dm@jxnu.edu.cn}
\affiliation{ School of Physics, Jiangxi Normal University, Nanchang 330022, China}
\author{Yan-Qing Ma}
\email{yqma@pku.edu.cn}
\affiliation{School of Physics and State Key Laboratory of Nuclear Physics and Technology, Peking University, Beijing 100871, China}
\affiliation{ Center for High Energy Physics, Peking University,
Beijing 100871, China}


\date{}

\begin{abstract}
Prompt $J/\psi$ hadroproduction poses two related puzzles: fixed-order hadronic and $e^+e^-$ analyses give incompatible constraints on the same color-octet~(CO) long-distance matrix element~(LDME) combination, and the weak observed polarization is difficult to reproduce together with the yield. We analyze high-$p_T$ production in the fragmentation-function formalism with leading-logarithmic threshold resummation in both nonrelativistic QCD~(NRQCD) and soft-gluon factorization~(SGF). Combining the resummed hadroproduction fit with bounds from resummed $e^+e^-$ production, we find in both frameworks a common LDME region satisfying the two constraints simultaneously: threshold resummation resolves the universality tension present at fixed order. There, the $\COaSz$ contribution is small and the yield is governed by the $\COcSa$ and $\COcPj$ channels, whose leading transverse contributions largely cancel. The cancellation amplifies spin-symmetry-breaking transitions in the $\COcSa$ channel, so that within a one-parameter spin-flip ansatz a contribution of natural order $v^4$ accommodates both the measured yields and the weak polarization. A similar sensitivity holds for $\psi(2S)$ production.
\end{abstract}

\maketitle
\allowdisplaybreaks

\sect{Introduction}
Nonrelativistic QCD~(NRQCD) factorization provides an effective-field-theory framework for heavy-quarkonium production and decay~\cite{Bodwin:1994jh}, separating perturbatively calculable short-distance coefficients from nonperturbative long-distance matrix elements~(LDMEs) organized in powers of the heavy-quark velocity $v$. It has enabled detailed calculations of quarkonium yields and polarization in a wide range of collision processes~\cite{Campbell:2007ws,Gong:2008sn,Butenschoen:2011yh,Ma:2010yw,Butenschoen:2010rq,Ma:2010jj,Ma:2010vd,Wang:2012is,Gong:2013qka,Bodwin:2014gia,Shao:2014yta,Han:2014kxa,Han:2014jya,Zhang:2014ybe,Zhang:2014coi,Bodwin:2015iua,Feng:2020cvm,Butenschoen:2022qka,Brambilla:2024iqg}. Nevertheless, no single color-octet~(CO) LDME set has yet provided a fully consistent description of prompt $J/\psi$ production.

Two related tensions persist. The first concerns LDME universality. At large $p_T$, the hadroproduction spectrum effectively constrains only two linear combinations of the CO LDMEs,
\begin{align}
M_{r_0}^{J/\psi}
&\equiv \langle\mathcal O^{J/\psi}(\COaSz)\rangle
+\frac{r_0}{m_c^2}\langle\mathcal O^{J/\psi}(\COcPz)\rangle,
\nonumber\\
M_{r_1}^{J/\psi}
&\equiv \langle\mathcal O^{J/\psi}(\COcSa)\rangle
+\frac{r_1}{m_c^2}\langle\mathcal O^{J/\psi}(\COcPz)\rangle,
\end{align}
with $m_c$ the charm-quark mass. In the fixed-order analysis of Ref.~\cite{Ma:2010yw}, $r_0=3.9$ and $r_1=-0.56$, with fitted values $M_{r_0}^{J/\psi}=0.074~\textrm{GeV}^3$ and $M_{r_1}^{J/\psi}=0.0005~\textrm{GeV}^3$. The fixed-order analysis of $e^+e^-\to J/\psi+X$, by contrast, gives $M_{r_0=3.9}^{J/\psi}<0.02~\textrm{GeV}^3$~\cite{Zhang:2009ym}, substantially below the hadroproduction preference: within the fixed-order treatments, the two processes admit no common value.

The second tension concerns polarization. Prompt $J/\psi$ mesons are observed to be only weakly polarized over a broad kinematic range~\cite{CDF:2007msx,ALICE:2011gej,LHCb:2013izl,CMS:2013gbz}, whereas the leading-order color-octet picture predicts a sizable transverse component at large $p_T$~\cite{Cho:1994ih,Braaten:1999qk}. Several NLO fits that reproduce the weak polarization favor a comparatively large $M_{r_0=3.9}^{J/\psi}$~\cite{Gong:2008hk,Butenschoen:2012px,Chao:2012iv,Gong:2012ug,Bodwin:2014gia,Brambilla:2024iqg}. The two tensions are thus connected through the same LDME direction: increasing $M_{r_0=3.9}^{J/\psi}$ improves the polarization description but simultaneously inflates the predicted $e^+e^-$ rate, violating the fixed-order bound. Within the fixed-order treatment, yield, polarization, and universality are therefore difficult to reconcile.

These tensions motivate examining perturbative effects beyond fixed order. In the fragmentation-function description, the matching coefficients of several quarkonium channels contain logarithmically enhanced distributions near the endpoint $z\to1$~\cite{Bauer:2001rh,Fleming:2003gt,Fleming:2006cd,Chen:2021hzo}, whose impact is amplified by the steeply falling partonic spectrum, making order-by-order expansions unreliable. Threshold resummation organizes these contributions to all orders and has been developed for quarkonium hadroproduction in both soft-gluon factorization~(SGF) and NRQCD~\cite{Chen:2021hzo,Chen:2023gsu,Chung:2024jfk}: it removes negative fixed-order predictions for high-$p_T$ $\chi_{cJ}$ and $J/\psi$ production~\cite{Chung:2023ext,Chen:2023gsu,Chung:2024jfk}, and it modifies the shapes and relative normalizations of the $\COcSa$ and $\COcPj$ contributions, thereby changing the effective projection coefficients $r_0$ and $r_1$. Separately, endpoint resummation in $e^+e^-\to J/\psi+X$ weakens the upper bound on the $\COaSz$-$\COcPz$ LDME combination~\cite{Chen:2022qli}. These results motivate a consistent test of whether the resummed hadroproduction fit overlaps the region allowed by resummed $e^+e^-$ production and what polarization pattern follows.

In this Letter, we perform this test by analyzing high-$p_T$ prompt $J/\psi$ hadroproduction with threshold resummation in both NRQCD and SGF, fitting the spectrum subject to the resummed $e^+e^-$ upper bound on the $\COaSz$-$\COcPz$ LDME combination. In both frameworks, the constrained fit admits a nonempty LDME region compatible with the cross-process bound, in which the $\COaSz$ contribution is small and the yield is dominated by the $\COcSa$ and $\COcPj$ channels, whose leading transverse components have opposite signs and largely cancel. We then include spin-symmetry-breaking transitions in the $\COcSa$ channel through a one-parameter spin-flip ansatz. Since no corresponding term is included for the $\COcPj$ channel, the transverse cancellation suppresses the spin-preserving rate while leaving the longitudinal spin-flip term untouched, promoting a formally $v^4$-suppressed effect to the leading polarization contribution, which accommodates the observed weak polarization without degrading the yield description. The Supplemental Material shows a similar sensitivity for $\psi(2S)$ and presents the $\chi_{cJ}$ and $\eta_c$ analyses used to constrain feeddown and related LDMEs.

\sect{Theoretical framework}\label{sec:framework}
At transverse momentum $p_T$ much larger than the heavy-quark mass $m_Q$, the inclusive cross section for producing a quarkonium state $H$ with momentum $p$ admits an expansion in powers of $m_Q^2/p_T^2$~\cite{Kang:2014tta,Kang:2014pya}:
\begin{align}\label{eq:pqcdfac}
\mathrm{d}\sigma_{A+B\to H+X}(p) \approx &
\sum_{i,j}f_{i/A}(x_1,\mu_F)f_{j/B}(x_2,\mu_F)
 \nonumber\\
&\hspace{-3cm}   \times \Big\{ \sum_{f} D_{f\to H}(z,\mu_F)
    \otimes
\mathrm{d}{\hat{\sigma}}_{i+j\to f+X}({\hat P}/z,\mu_F)
 \nonumber\\
&\hspace{-3cm} +  \sum_{\kappa}  {\cal
D}_{[Q\bar{Q}(\kappa)]\to H}(z,\zeta,\zeta',\mu_F)
  \\
&\hspace{-3cm} \otimes
\mathrm{d}{\hat{\sigma}}_{i+j\to [Q\bar{Q}(\kappa)]+X}({\hat P}(1\pm\zeta)/2z,{\hat
P}(1\pm\zeta')/2z,\mu_F)   \Big\}, \nonumber
\end{align}
Here $\sum_f$ runs over the fragmenting parton species and $\sum_\kappa$ over the spin-color states $\kappa$ of the fragmenting $Q\bar Q$ pair. The functions $f_{i/A}(x_1,\mu_F)$ and $f_{j/B}(x_2,\mu_F)$ are the parton distribution functions of the incoming hadrons $A$ and $B$, with $x_1$ and $x_2$ the corresponding longitudinal momentum fractions, and $\mu_F$ is the collinear factorization scale. The symbol $\otimes$ denotes convolution in the momentum fractions on each side of it. The first term in braces is the leading-power~(LP) contribution from single-parton fragmentation and scales as $p_T^{-4}$; the second is the heavy-quark-pair fragmentation contribution retained at next-to-leading power~(NLP), which scales as $p_T^{-6}$~\cite{Kang:2014tta,Kang:2014pya}. The quarkonium carries momentum $p$, while $\hat P^\mu=(p^+,0,\boldsymbol{0}_\perp)$ is a lightlike vector with the same plus component as $p^\mu$. The variable $z$ is the fraction of the fragmenting system's plus momentum carried by $H$, while $\zeta$ and $\zeta'$ specify the relative heavy-quark momenta in the amplitude and its conjugate, respectively; the corresponding quark and antiquark momenta in the second hard part are $\hat P(1\pm\zeta)/(2z)$ and $\hat P(1\pm\zeta')/(2z)$.

The fragmentation functions~(FFs) are initialized at a perturbative scale $\mu_0\gtrsim2m_Q$ and then evolved to $\mu_F$. Their nonperturbative content can be organized either in NRQCD or in SGF. For compactness, we write $\kappa$ in a FF or matching coefficient as shorthand for the pair state $[Q\bar Q(\kappa)]$. In NRQCD, after truncating the velocity expansion, the input FFs are matched onto LDMEs as~\cite{Bodwin:1994jh}
\begin{subequations}\label{eq:NRQCD}
\begin{align}
D_{f\to H}(z,\mu_0)
={}&\sum_n\hat d_{f\to n}(z,\mu_0,\mu_\Lambda)
\langle\mathcal O^H(n)\rangle,\\
{\cal D}_{\kappa\to H}(z,\zeta,\zeta',\mu_0)
={}&\sum_n\hat d_{\kappa\to n}
(z,\zeta,\zeta',\mu_0,\mu_\Lambda)
\langle\mathcal O^H(n)\rangle.
\end{align}
\end{subequations}
The label $n={}^{2S+1}L_J^{[c]}$ specifies the angular-momentum and color quantum numbers of the intermediate pair, with $c=1$ or $8$. The coefficient $\hat d$ is a perturbative NRQCD matching coefficient for the FF, $\langle\mathcal O^H(n)\rangle\equiv\langle\mathcal O^H(n)\rangle(\mu_\Lambda)$ is the corresponding LDME, and $\mu_\Lambda$ is the NRQCD factorization scale. For the channels specified below, we resum the leading threshold logarithms in $\hat d$ at leading-logarithmic~(LL) accuracy by Mellin-space exponentiation and match the result to the full lowest-nonvanishing-order coefficient~\cite{Chung:2024jfk}. This resummation changes both the shapes and relative normalizations of the CO contributions, particularly the $\COcSa$ and $\COcPj$ channels.

In SGF, the same input FFs are organized as convolutions of perturbative matching coefficients with soft-gluon distributions~(SGDs) $F_{[n]\to H}$, which describe the transition $Q\bar Q[n]\to H+X$ at the SGF factorization scale $\mu_f$. The LL soft-enhanced terms are assigned to the SGD and resummed through its renormalization-group evolution, so that the two frameworks share the same large-$p_T$ factorization formula but organize the endpoint dynamics into different nonperturbative quantities. The corresponding expressions are given in the Supplemental Material~\cite{Ma:2017xno,Chen:2021hzo,Chen:2023gsu,Jia:2024cvv}.

For the calculation below, we retain gluon fragmentation, $f=g$, in the LP term of Eq.~\eqref{eq:pqcdfac} and the heavy-quark-pair fragmentation contribution at NLP. The gluon FF starts at $\mathcal O(\alpha_s)$ for $\COcSa$ and at $\mathcal O(\alpha_s^2)$ for $\COaSz$, $\COcPj$, and $\CScPj$; the nonvanishing pair FFs relevant to the retained channels start at $\mathcal O(\alpha_s^0)$. We combine these lowest-nonvanishing-order inputs with LL threshold resummation. Direct $\CScSa$ production is omitted because it remains below the estimated theoretical uncertainty in the kinematic region considered~\cite{Bodwin:2015iua}. The matching coefficients, resummation formulas, SGD evolution, and model inputs are given in the Supplemental Material.

\sect{Universality of LDMEs}
We now turn to the phenomenology. The calculation combines LO PDFs and LO hard parts with the LL-resummed FFs introduced above in Eq.~\eqref{eq:pqcdfac}. We adopt the CTEQ6L1 PDF set~\cite{Pumplin:2002vw} with LO running of $\alpha_s$ for $n_f=5$ and $\Lambda_{\rm QCD}^{(5)}=165~\rm MeV$; the LP hard parts at $\mathcal O(\alpha_s^2)$ are taken from Ref.~\cite{Aversa:1988vb} and the NLP hard parts at $\mathcal O(\alpha_s^3)$ from Refs.~\cite{Kang:2011mg,Kang:2014pya}.

We fix the charm-quark mass to $m_c=1.5~\rm GeV$ and set $\mu_F=p_T$ in Eq.~\eqref{eq:pqcdfac}. For the nonperturbative inputs, we take $\mu_\Lambda=m_c$ in NRQCD and $\mu_f=M_H/2$ in SGF. The single-parton FFs are evolved from $\mu_0=2m_c$ in NRQCD and $\mu_0=M_H$ in SGF up to $\mu_F=p_T$, using the method of Ref.~\cite{Bodwin:2015iua} with the LO kernel for $n_f=3$, thereby resumming the logarithms of $p_T^2/m_c^2$.\footnote{We do not evolve the double-parton FFs because the corresponding renormalization-group solution is not yet available. This omission is expected to be numerically acceptable: the large logarithms matter mainly at high $p_T$, where the NLP contribution is already suppressed relative to single-parton fragmentation.} We use $M_{\eta_c}=3~\rm GeV$, $M_{\chi_{c1}}=3.511~\rm GeV$, $M_{\chi_{c2}}=3.556~\rm GeV$, $M_{\psi(2S)}=3.686~\rm GeV$, and $M_{J/\psi}=3.1~\rm GeV$~\cite{ParticleDataGroup:2010dbb}. For feeddown from $H\to J/\psi+X$, we approximate $p_T^H=(M_H/M_{J/\psi})p_T^{J/\psi}$~\cite{Ma:2010yw} and take the relevant branching ratios from Ref.~\cite{Shao:2014yta}.

To determine the LDMEs, we combine the CMS $J/\psi$ transverse-momentum spectrum~\cite{CMS:2015lbl} at $\sqrt s=7~\rm TeV$ and $|y|<1.2$ with the upper bounds on $M_{r_0}^{J/\psi}$ obtained from resummed $e^+e^-$ production~\cite{Chen:2022qli}:
\begin{subequations}\label{eq:ee-M0}
\begin{align}
M_{r_0=3.9}^{J/\psi}\vert_{\textrm{NRQCD}}
& < 7.6 \times 10^{-2} \ \textrm{GeV}^3, \\
M_{r_0=2.5}^{J/\psi}\vert_{\textrm{SGF}}
& < 9.4 \times 10^{-2} \ \textrm{GeV}^3.
\end{align}
\end{subequations}
The fit is therefore required not only to describe the hadroproduction data but also to remain compatible with this cross-process constraint. Because Eq.~\eqref{eq:pqcdfac} is reliable only at large $p_T$, only data with $p_T\ge 12~\rm GeV$ are included. The theoretical cross sections are assigned a $30\%$ uncertainty to account for missing relativistic corrections~\cite{Bodwin:2015iua}.

The direct CO contribution to the $J/\psi$ cross section then decomposes as
\begin{align}\label{eq:decom}
d\sigma^{J/\psi} =& d\hat{\sigma}[\COcSa] \langle\mathcal{O}^{J/\psi}(\COcSa)\rangle + d\hat{\sigma}[\COaSz] \langle\mathcal{O}^{J/\psi}(\COaSz)\rangle \nonumber\\
 & + d\hat{\sigma}[\COcPz] \langle\mathcal{O}^{J/\psi}(\COcPz)\rangle,
\end{align}
where $d\hat{\sigma}[n]$ denotes the resummed contribution of channel $n$ in Eq.~\eqref{eq:pqcdfac}, including the PDFs, hard parts, and evolved FFs, and the feeddown contributions are added as described above. Consistent with earlier analyses~\cite{Ma:2010yw,Ma:2010jj,Shao:2014yta}, we find that, over the fitted region $12~\rm GeV\le p_T\le 120~\rm GeV$ and $|y|<1.2$, the $\COcPz$ short-distance contribution is well approximated by a linear projection onto the $\COaSz$ and $\COcSa$ coefficients:
\begin{align}\label{eq:decom-sdc}
d\hat{\sigma}[\COcPz]\vert_X = r_0^X \frac{d\hat{\sigma}[\COaSz]}{m_c^2} + r_1^X \frac{d\hat{\sigma}[\COcSa]}{m_c^2},
\end{align}
where $X\in\{\textrm{NRQCD},\textrm{SGF}\}$ labels the factorization framework. The resummed projection coefficients, $r_0^{\textrm{NRQCD}}=24.45$ and $r_1^{\textrm{NRQCD}}=-5.34$ in NRQCD and $r_0^{\textrm{SGF}}=6.49$ and $r_1^{\textrm{SGF}}=-3.41$ in SGF, differ markedly from the fixed-order values $r_0=3.9$ and $r_1=-0.56$ quoted in the Introduction: threshold resummation reshuffles the relative importance of the $\COcSa$ and $\COcPj$ channels. This enlarged hierarchy opens enough LDME parameter space for the hadroproduction fit to satisfy the cross-process constraints of Eq.~\eqref{eq:ee-M0}.
Performing a minimum-$\chi^2$ fit to the CMS data~\cite{CMS:2015lbl} and imposing Eq.~\eqref{eq:ee-M0}, we obtain the allowed intervals
\begin{subequations}\label{eq:M00}
\begin{align}
2.8 \times 10^{-2}< M_{r_0=3.9}^{J/\psi}\vert_{\textrm{NRQCD}}&
 < 7.6 \times 10^{-2} \ \textrm{GeV}^3, \\
5.4 \times 10^{-2}<M_{r_0=2.5}^{J/\psi}\vert_{\textrm{SGF}}&
 < 9.4 \times 10^{-2} \ \textrm{GeV}^3.
\end{align}
\end{subequations}
The corresponding ranges for the individual LDMEs, obtained with the additional requirement that all CO LDMEs be positive~\cite{Shao:2014yta}, are
\begin{subequations}\label{eq:NRQCD-constraint}
\begin{align}
3.0\times 10^{-2} < \langle\mathcal{O}^{J/\psi}(\COcSa)\rangle& < 8.8 \times 10^{-2} \ \textrm{GeV}^3,\\
0 < \langle\mathcal{O}^{J/\psi}(\COaSz)\rangle& < 5.7 \times 10^{-2} \ \textrm{GeV}^3,\\
0.5 \times 10^{-2} < \frac{\langle\mathcal{O}^{J/\psi}(\COcPz)\rangle}{m_c^2}& < 1.9 \times 10^{-2} \ \textrm{GeV}^3,
\end{align}
\end{subequations}
in NRQCD factorization, and
\begin{subequations}\label{eq:SGF-constraint}
\begin{align}
9.1\times 10^{-2} < \langle\mathcal{O}^{J/\psi}(\COcSa)\rangle& < 17.9 \times 10^{-2} \ \textrm{GeV}^3,\\
0 < \langle\mathcal{O}^{J/\psi}(\COaSz)\rangle& < 6.6 \times 10^{-2} \ \textrm{GeV}^3,\\
1.1 \times 10^{-2} < \frac{\langle\mathcal{O}^{J/\psi}(\COcPz)\rangle}{m_c^2}& < 3.8 \times 10^{-2} \ \textrm{GeV}^3,
\end{align}
\end{subequations}
in SGF. The hadroproduction data alone constrain only two combinations of the three CO LDMEs; the individual ranges above are shaped jointly by the cross-process bound and the positivity requirement, and a nonempty allowed region common to both constraints remains.

\begin{figure*}
\begin{center}
\begin{tabular}{ccc}
\includegraphics[width=0.3\textwidth]{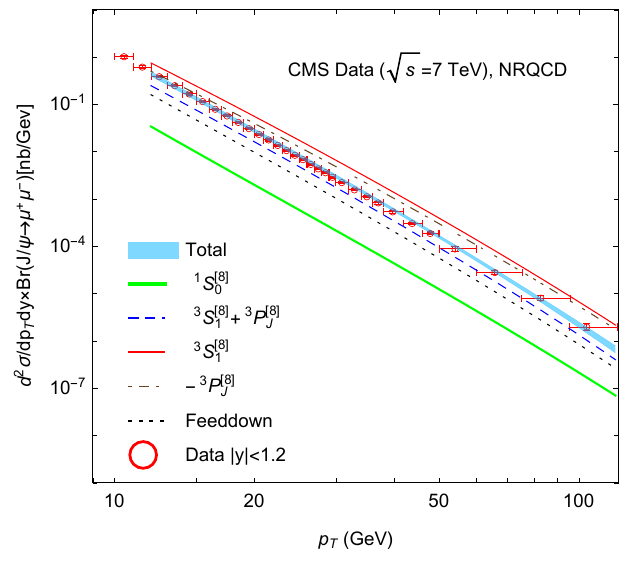}
&
\includegraphics[width=0.3\textwidth]{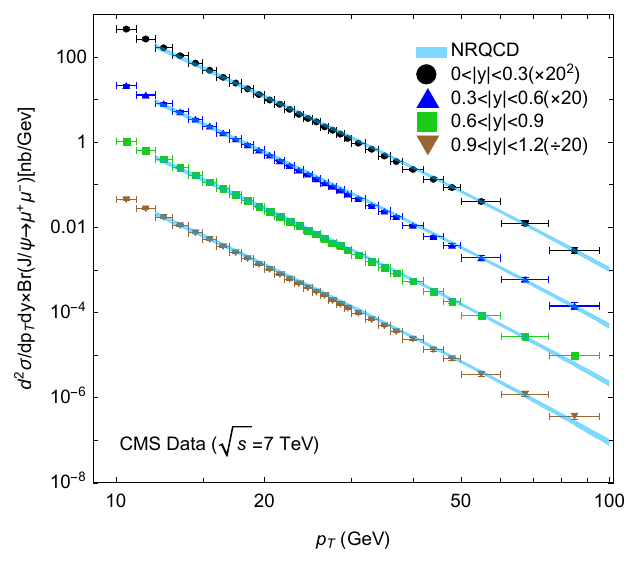}
&
\includegraphics[width=0.3\textwidth]{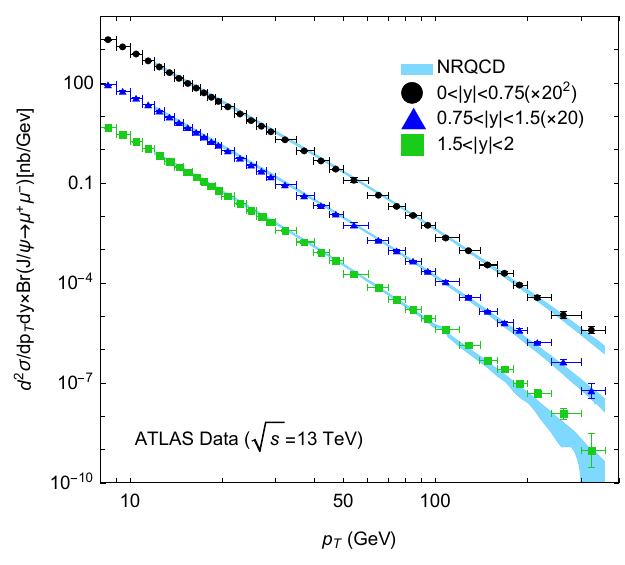}
\\
\includegraphics[width=0.3\textwidth]{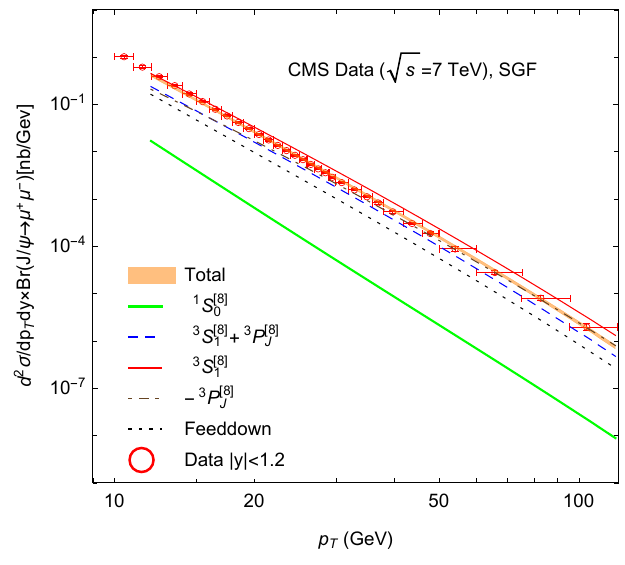}
&
\includegraphics[width=0.3\textwidth]{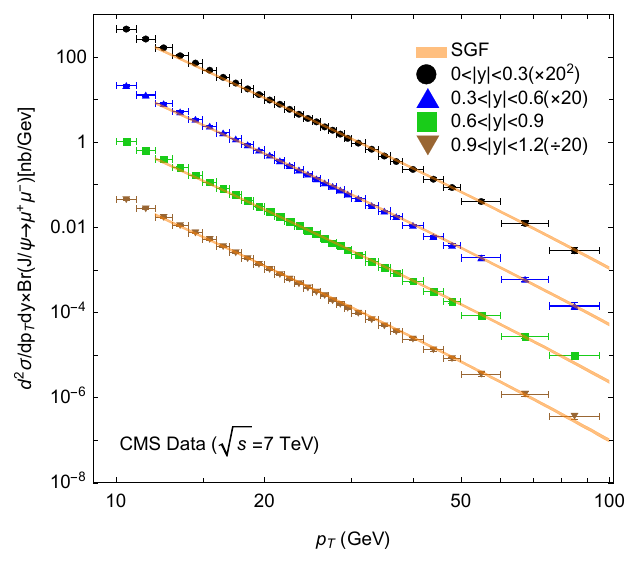}
&
\includegraphics[width=0.3\textwidth]{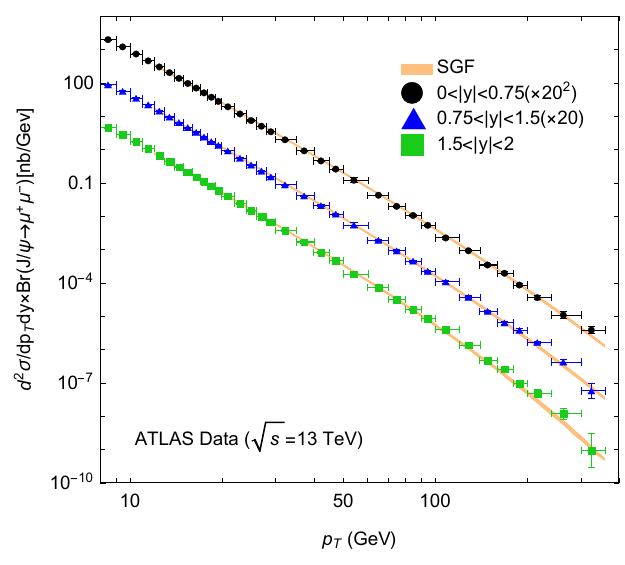}
\end{tabular}
\end{center}
\caption{Upper (lower) row: results in NRQCD (SGF). Left column: fitted prompt-$J/\psi$ $p_T$ spectra at $\sqrt{s}=7~\rm TeV$ and $|y|<1.2$, with the individual channel contributions. Middle and right columns: predictions based on the fitted LDMEs, compared with CMS data at $\sqrt{s}=7$ and ATLAS data at $13~\rm TeV$, respectively. $\mathrm{Br}$ denotes the branching ratios.}\label{fig:ptdata}
\end{figure*}

Figure~\ref{fig:ptdata} shows the resulting description of the yield data. The left column displays the fitted $p_T$ spectrum at $\sqrt s=7~\rm TeV$ and $|y|<1.2$, decomposed into the individual channel contributions. Once universality is imposed, the $\COaSz$ contribution is small, consistent with the independent bound from $\eta_c$ hadroproduction discussed in the Supplemental Material. Unlike the fits of Refs.~\cite{Chao:2012iv,Bodwin:2014gia,Bodwin:2015iua,Faccioli:2014cqa}, the yield is not dominated by $\COaSz$; it is controlled by the combined $\COcSa$ and $\COcPj$ channels. This altered channel hierarchy is what enables the polarization mechanism discussed in the next section.
The middle and right columns of Fig.~\ref{fig:ptdata} compare these predictions with CMS measurements at $\sqrt s=7$ and ATLAS measurements at $13~\rm TeV$ in several rapidity intervals. The agreement is satisfactory in both frameworks. Within the present resummed setup, the $pp$ fit and the $e^+e^-$ bound are therefore satisfied simultaneously, without degrading the description of the yield data.

\sect{Spin-flip effects and polarization}
We next turn to polarization, where the channel hierarchy implied by the resummed fit becomes decisive. Throughout this section, $d\sigma_T\equiv d\sigma_{+1}+d\sigma_{-1}$ and $d\sigma_L\equiv d\sigma_0$ denote the transverse and longitudinal cross sections in the $J/\psi$ helicity frame, and the polarization parameter is $\lambda_\theta=(d\sigma_T-2d\sigma_L)/(d\sigma_T+2d\sigma_L)$. Keeping the channels that dominate after the resummed fit, the direct polarized cross section reads
\begin{align}\label{eq:polar}
&d\sigma^{J/\psi}_{\lambda} =   d\hat{\sigma}[\COaSz] \langle\mathcal{O}^{J/\psi(\lambda)}(\COaSz) \rangle+\sum_{\lambda^\prime=T,L} \Big[d\hat{\sigma}[\state{{3}}{S}{1,\lambda^\prime}{8}]
\nonumber\\
 & \times \langle\mathcal{O}^{J/\psi(\lambda)}(\state{{3}}{S}{1,\lambda^\prime}{8})\rangle  + \frac{d\hat{\sigma}[\state{{3,\lambda^\prime}}{P}{}{8}]}{m_c^2} \langle\mathcal{O}^{J/\psi(\lambda)}(\state{{3,\lambda^\prime}}{P}{}{8})\rangle\Big].
\end{align}
Here $\mathcal O^{J/\psi(\lambda)}(n)$ is the LDME for producing a $J/\psi$ with helicity $\lambda$ through the intermediate state $n$, the labels $T$ and $L$ on the intermediate states denote the summed transverse ($\lambda'=\pm1$) and longitudinal ($\lambda'=0$) polarizations of the $Q\bar Q$ pair, and the aggregated $P$-wave state $\state{{3,\lambda^\prime}}{P}{}{8}$ sums over $J=0,1,2$. Our calculation yields $d\hat{\sigma}[\state{{3}}{S}{1,L}{8}]\ll d\hat{\sigma}[\state{{3}}{S}{1,T}{8}]$ and $d\hat{\sigma}[\state{{3,L}}{P}{}{8}]\ll d\hat{\sigma}[\state{{3,T}}{P}{}{8}]$: both CO channels produce predominantly transverse pairs. Transitions that break heavy-quark spin symmetry are suppressed by $v^4\approx 0.09$ (with $v^2\approx 0.3$ for the $J/\psi$) relative to the spin-symmetric pieces~\cite{Cho:1994ih,Beneke:1995yb} and are often neglected; with the $\COaSz$ contribution small, one would then expect $d\sigma_T^{J/\psi}\gg d\sigma_L^{J/\psi}$, the strongly transverse polarization that constitutes the conventional puzzle. The resummed fit, however, selects a different hierarchy: the yield is controlled by the $\COcSa$ and $\COcPj$ channels, whose leading transverse contributions largely cancel. In the one-parameter description below, spin-symmetry-breaking transitions are retained only in the $\COcSa$ channel: with the longitudinal pair channels negligible, the spin-flip transition of the transverse $\state{{3}}{S}{1,T}{8}$ state is the leading source of longitudinal $J/\psi$ production, whereas spin-flip matrix elements in the other channels would introduce additional independent parameters and are neglected. The transverse cancellation then acts as an amplifier: it suppresses the spin-preserving rate while leaving the spin-flip residue untouched, so that a formally $v^4$-suppressed effect is promoted to a leading contribution to the polarization.

To make this explicit, we parametrize the spin-flip contribution in the $\COcSa$ channel as
\begin{subequations}\label{eq:spin-flip}
\begin{align}
\langle\mathcal{O}^{J/\psi(L)}(\state{{3}}{S}{1,T}{8})\rangle=&\frac{\xi}{1+\xi}\frac{1}{3}\langle \mathcal{O}^{J/\psi}(\COcSa)\rangle,\\
\langle\mathcal{O}^{J/\psi(T)}(\state{{3}}{S}{1,T}{8})\rangle=&\frac{1}{1+\xi}\frac{1}{3}\langle \mathcal{O}^{J/\psi}(\COcSa)\rangle,
\end{align}
\end{subequations}
where a transversely polarized $\state{{3}}{S}{1,T}{8}$ pair produces a transverse (longitudinal) $J/\psi$ with relative weight $1/(1+\xi)$ ($\xi/(1+\xi)$), so that the two matrix elements sum to the spin-averaged value $\langle\mathcal O^{J/\psi}(\COcSa)\rangle/3$ and the LDME fitted in the previous section is left unchanged. The parameter $\xi$ is expected to be of order $v^4$. Dropping the small $\COaSz$ term and the longitudinal pair channels in Eq.~\eqref{eq:polar}, and relating the polarized $P$-wave matrix elements to $\langle \mathcal{O}^{J/\psi}(\COcPz)\rangle$ by heavy-quark spin symmetry, Eqs.~\eqref{eq:polar} and~\eqref{eq:spin-flip} yield
\begin{subequations}\label{eq:T-L}
\begin{align}
d\sigma^{J/\psi}_{T} \approx &    \frac{1}{3(1+\xi)}d\hat{\sigma}[\state{{3}}{S}{1,T}{8}]
\langle \mathcal{O}^{J/\psi}(\COcSa)\rangle  \nonumber\\
&+ \frac{d\hat{\sigma}[\state{{3,T}}{P}{}{8}]}{m_c^2} \langle \mathcal{O}^{J/\psi}(\COcPz)\rangle,\\
d\sigma^{J/\psi}_{L}  \approx &  \frac{\xi}{3(1+\xi)} d\hat{\sigma}[\state{{3}}{S}{1,T}{8}]
\langle \mathcal{O}^{J/\psi}(\COcSa)\rangle.
\end{align}
\end{subequations}
The first equation in Eq.~\eqref{eq:T-L} exhibits the key cancellation: its two terms, the transverse $\COcSa$ and $P$-wave contributions, have opposite signs and largely cancel. The second equation shows that the longitudinal rate is generated by the spin-flip residue of the $\COcSa$ channel. Since the $\COcPj$ channel carries no corresponding spin-flip term, the cancellation in $d\sigma_T^{J/\psi}$ is not accompanied by one in $d\sigma_L^{J/\psi}$: $d\sigma_L^{J/\psi}$ becomes comparable to the surviving $d\sigma_T^{J/\psi}$ even though $\xi$ is only of order $v^4$. The sensitivity to spin-flip effects thus reflects the transverse cancellation rather than an anomalously large spin-flip matrix element.

\begin{figure*}
\begin{center}
\begin{tabular}{cccc}
\includegraphics[width=0.25\textwidth]{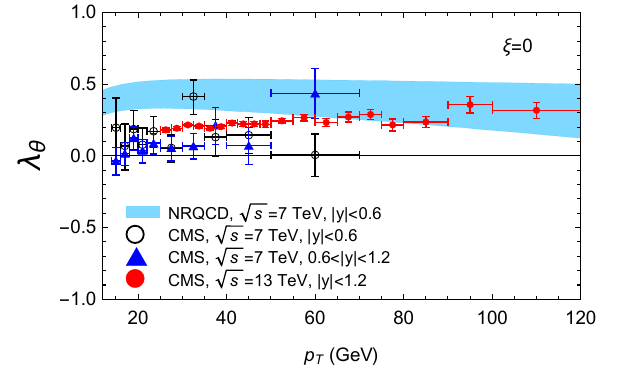}
&
\includegraphics[width=0.25\textwidth]{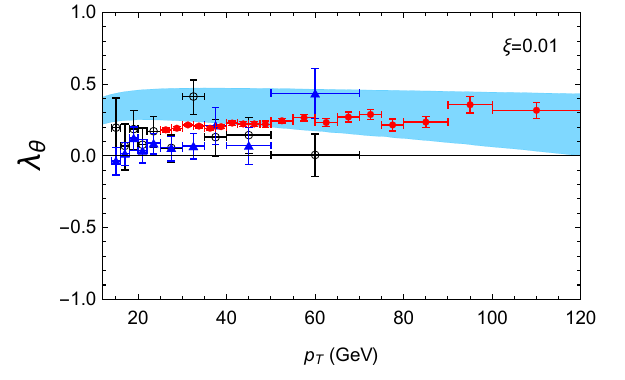}
&
\includegraphics[width=0.25\textwidth]{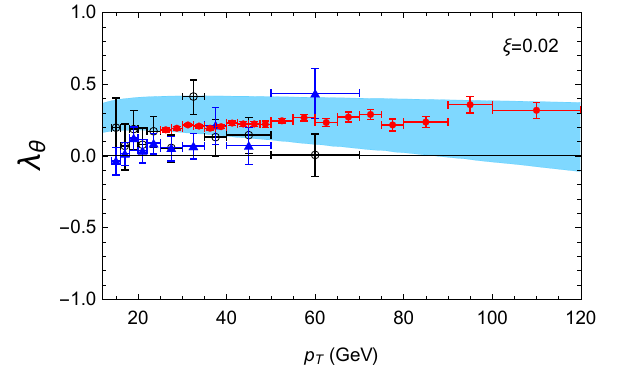}
&
\includegraphics[width=0.25\textwidth]{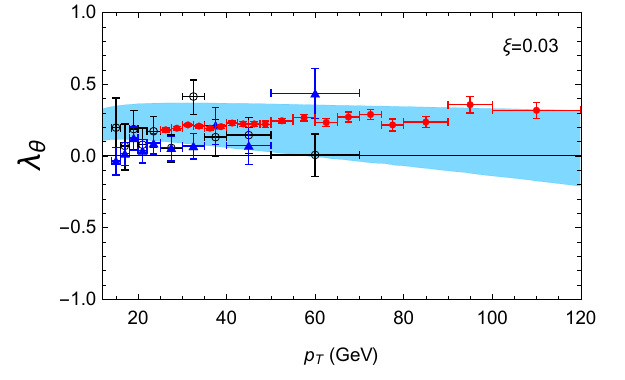}
\\
\includegraphics[width=0.25\textwidth]{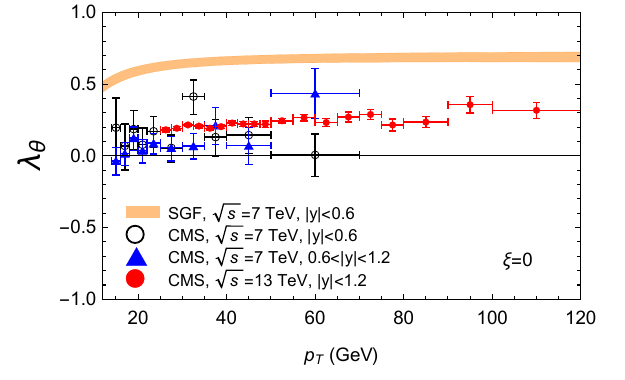}
&
\includegraphics[width=0.25\textwidth]{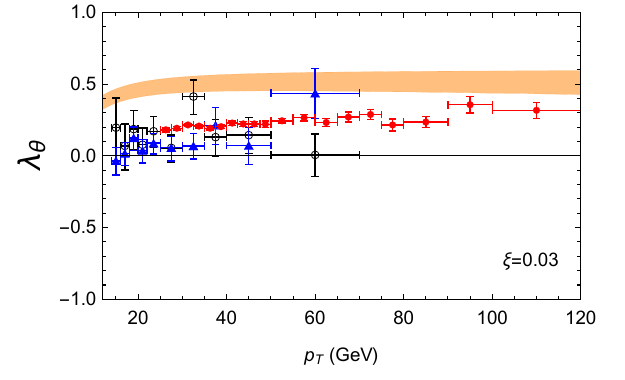}
&
\includegraphics[width=0.25\textwidth]{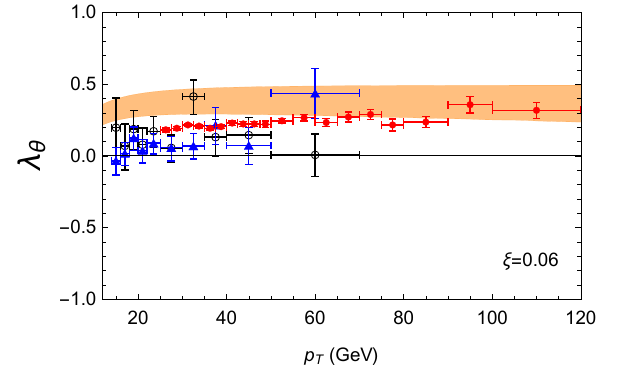}
&
\includegraphics[width=0.25\textwidth]{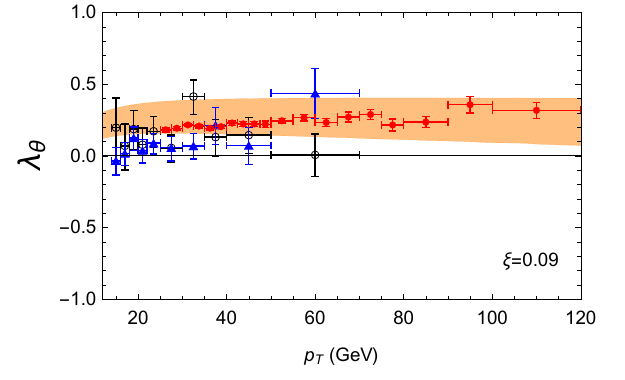}
\end{tabular}
\end{center}
\caption{Upper (lower) row: results in NRQCD (SGF). The four columns correspond, from left to right, to four representative values of $\xi$. The curves show the predicted prompt-$J/\psi$ polarization parameter $\lambda_\theta=(d\sigma_T-2d\sigma_L)/(d\sigma_T+2d\sigma_L)$ in the helicity frame, compared with the CMS data~\cite{CMS:2013gbz,CMS:2024igk}. }\label{fig:cms}
\end{figure*}

Figure~\ref{fig:cms} shows $\lambda_\theta$ as a function of $p_T$ for four representative values of $\xi$, in both NRQCD and SGF, compared with the CMS polarization measurements~\cite{CMS:2013gbz,CMS:2024igk}. As $\xi$ increases from zero, $\lambda_\theta$ moves away from the transverse limit, and for values of order $v^4$ the predictions agree well with the data, particularly for the $13~\rm TeV$ measurements, which reach $\lambda_\theta\sim 0.3$ at large $p_T$. The observed nearly unpolarized state is thus a direct consequence of the channel hierarchy selected by the resummed, cross-process-consistent fit.

\sect{Summary}
We have studied high-$p_T$ prompt $J/\psi$ hadroproduction in the fragmentation-function formalism with leading-logarithmic threshold resummation, implemented in both NRQCD and SGF. Since the resummed $\COcPj$ coefficient is well approximated by a linear projection onto the $\COaSz$ and $\COcSa$ coefficients, the hadroproduction data constrain only two combinations of the CO LDMEs. Combining the resummed hadroproduction fit with the upper bounds from resummed $e^+e^-$ production, we find in both frameworks a nonempty LDME region satisfying the cross-process constraint, in which the $\COaSz$ contribution is small and the yield is dominated by the $\COcSa$ and $\COcPj$ channels. The universality tension reported in fixed-order analyses is thus resolved within the resummed setup.

The same channel hierarchy clarifies the polarization puzzle. The leading transverse $\COcSa$ and $\COcPj$ contributions largely cancel, and in our one-parameter treatment the spin-flip residue of the $\COcSa$ channel is not subject to this cancellation. The cancellation thereby amplifies the relative size of the spin-flip contribution, enabling a term of natural size $\mathcal O(v^4)$ to drive the prediction toward the observed weak polarization without degrading the yield description. The Supplemental Material shows an analogous sensitivity for $\psi(2S)$, together with auxiliary constraints on feeddown and LDMEs from the $\chi_{cJ}$ and $\eta_c$ analyses.

\sect{Acknowledgments}
The work of A.-P. Chen is supported by the National Natural Science Foundation of China (Nos.~12565013, 12205124) and the Jiangxi Provincial Natural Science Foundation (Grant No.~20242BAB20034). J. Guo is supported by the National Natural Science Foundation of China (No.~12305111) and the Jiangxi Provincial Natural Science Foundation (Grant No.~20252BAC20016). Y.-Q. Ma is supported by the National Natural Science Foundation of China (No.~12325503) and the High-performance Computing Platform of Peking University.

\providecommand{\href}[2]{#2}\begingroup\raggedright\endgroup

\appendix
\begin{widetext}

\section[Supplemental Material for ``Resolving the universality and polarization puzzles in J/psi hadroproduction'']{Supplemental Material for ``Resolving the universality and polarization puzzles in $J/\psi$ hadroproduction"}

\subsection{Resummation of threshold logarithms in NRQCD factorization}
This subsection summarizes the ingredients needed for threshold resummation in NRQCD factorization and fixes the notation used in the main text.

In NRQCD factorization, the short-distance coefficients~(SDCs) of FFs admit the following singular expansion near threshold, $z=1$~\cite{Ma:2013yla,Ma:2014eja,Laenen:2010uz,Beneke:2018gvs}:
\begin{subequations}
\begin{align}
\hat{d}_{ f \to n }(z)=&\sum_{i=0}^\infty \alpha_s^{i+1}\Big[c_i\delta(1-z)+\sum_{j=0}^{2i-1}\Big(c_{ij}
\Big[\frac{\ln^j(1-z)}{1-z}\Big]_++d_{ij}\ln^j(1-z)\Big) +\mathcal{O}(1-z)\Big], \\
\hat{d}_{ \kappa \to n }(z, \zeta,\zeta^\prime)=&\sum_{i=0}^\infty \alpha_s^{i}\Big[c_i\delta(1-z)+\sum_{j=0}^{2i-1}\Big(c_{ij}
\Big[\frac{\ln^j(1-z)}{1-z}\Big]_++d_{ij}\ln^j(1-z)\Big) +\mathcal{O}(1-z)\Big].
\end{align}
\end{subequations}
In these expressions, the coefficients $c_i$, $c_{ij}$, and $d_{ij}$ depend on the parton and the channel $n$, and the sum over $j$ is empty for $i=0$. The terms with coefficients $c_i$ and $c_{ij}$ are the leading-power singular contributions, which we denote by $\hat{d}_{ f \to n }(z)\vert_{\textrm{endpoint}}$ and $\hat{d}_{ \kappa \to n }(z, \zeta,\zeta^\prime)\vert_{\textrm{endpoint}}$. The coefficients $c_{i,\,2i-1}$ for $i\geq 1$---and, for the $\state{3}{P}{J}{1,8}$ gluon FFs, $c_{i,\,2i-2}$ for $i\geq 2$---generate the leading-logarithmic~(LL) tower, while the $d_{ij}\ln^j(1-z)$ terms are less singular and are not resummed here. As $z\to 1$, the endpoint logarithms spoil the reliability of the fixed-order expansion and can even drive the cross section negative~\cite{Chen:2023gsu,Chung:2024jfk}.

For large-$p_T$ $J/\psi$ hadroproduction, the LP term in Eq.~\eqref{eq:pqcdfac} is dominated by gluon fragmentation ($f=g$). Threshold logarithms appear in the gluon FFs for the $\COcSa$, $\COcPj$, and $\CScPj$ channels~\cite{Ma:2013yla,Ma:2014eja,Zhang:2020atv}. They also arise in the double-parton channels $v^{[8]} \to \COcSa$, $a^{[8]} \to \COaSz$, $v^{[8]} \to \COcPz$, $a^{[8]} \to \COcPa$, and $v^{[8]} \to \COcPb$~\cite{Ma:2013yla,Ma:2014eja}, where $v^{[8]}$ and $a^{[8]}$ denote the color-octet vector and axial-vector pair states. Following Ref.~\cite{Chung:2024jfk}, we resum the LL terms in $\hat{d}_{ g \to n }$ in Mellin space. The Mellin transform and its inverse are
\begin{align}
\tilde{f}(N)=\int_0^1 \md z\, z^{N-1} f(z), \qquad
f(z)=\frac{1}{2\pi \mi}\int_{c-\mi\infty}^{c+\mi\infty} \md N\, z^{-N} \tilde{f}(N),
\end{align}
where the real constant $c$ is chosen to lie to the right of the rightmost singularity of $\tilde f(N)$. The LL contributions are then resummed by exponentiation~\cite{Chung:2024jfk}:
\begin{align}\label{eq:resum-endpoint-LP}
\tilde{\hat{d}}^{\textrm{resum}}_{ g \to n }(N)\Big\vert_{\textrm{endpoint}}=&\exp[J^{\textrm{LP}}_n(N)]\times \tilde{\hat{d}}^{\textrm{LO}}_{ g \to n }(N)\Big\vert_{\textrm{endpoint}},
\end{align}
with
\begin{subequations}
\begin{align}
J^{\textrm{LP}}_{\COcSa}(N)=&\frac{ \alpha_s C_A}{\pi} \int_0^1 dz z^{N-1}\Big[\frac{-2\ln(1-z)}{1-z}\Big]_+,\\
J^{\textrm{LP}}_{\COcPj}(N)=&J^{\textrm{LP}}_{\CScPj}(N)
=\frac{4}{3}J^{\textrm{LP}}_{\COcSa}(N).
\end{align}
\end{subequations}
The label ``LO'' in Eq.~\eqref{eq:resum-endpoint-LP} refers to the lowest nonvanishing order of $\tilde{\hat{d}}_{ g \to n }(N)\vert_{\textrm{endpoint}}$ in $\alpha_s$. It is of order $\alpha_s$ for $n=\COcSa$ and of order $\alpha_s^2$ for $n=\COcPj$ and $\CScPj$.

The LL terms in the double-parton FFs can be resummed in the same way:
\begin{align}\label{eq:resum-endpoint-NLP}
\tilde{\hat{d}}^{\textrm{resum}}_{ \kappa \to n }(N,\zeta,\zeta^\prime)\Big\vert_{\textrm{endpoint}}=&\exp[J^{\textrm{NLP}}_{
\kappa \to n}(N)] \times \tilde{\hat{d}}^{\textrm{LO}}_{ \kappa \to n }(N,\zeta,\zeta^\prime)\Big\vert_{\textrm{endpoint}},
\end{align}
where
\begin{align}
J^{\textrm{NLP}}_{v^{[8]}\to \COcSa}(N)
=&J^{\textrm{NLP}}_{a^{[8]}\to\COaSz}(N)=J^{\textrm{NLP}}_{v^{[8]}\to\COcPz}(N)
=J^{\textrm{NLP}}_{a^{[8]}\to\COcPa}(N)=J^{\textrm{NLP}}_{v^{[8]}\to\COcPb}(N)
=J^{\textrm{LP}}_{\COcSa}(N).
\end{align}
Combining Eqs.~\eqref{eq:resum-endpoint-LP} and~\eqref{eq:resum-endpoint-NLP}, we obtain the resummed SDCs:
\begin{subequations}\label{eq:resum-endpoint}
\begin{align}
\tilde{\hat{d}}^{\textrm{resum}}_{ g \to n }(N)
=&\,
\tilde{\hat{d}}_{ g \to n }(N)
+\tilde{\hat{d}}^{\textrm{resum}}_{ g \to n }(N)
\Big\vert_{\textrm{endpoint}}
-\Big[\tilde{\hat{d}}^{\textrm{resum}}_{ g \to n }(N)
\Big\vert_{\textrm{endpoint}}\Big]_{\textrm{FO}},
\\
\tilde{\hat{d}}^{\textrm{resum}}_{ \kappa \to n }
(N,\zeta,\zeta^\prime)
=&\,
\tilde{\hat{d}}_{ \kappa \to n }
(N,\zeta,\zeta^\prime)
+\tilde{\hat{d}}^{\textrm{resum}}_{ \kappa \to n }
(N,\zeta,\zeta^\prime)
\Big\vert_{\textrm{endpoint}}
-
\Big[\tilde{\hat{d}}^{\textrm{resum}}_{ \kappa \to n }
(N,\zeta,\zeta^\prime)
\Big\vert_{\textrm{endpoint}}\Big]_{\textrm{FO}} .
\end{align}
\end{subequations}
Here $[\cdots]_{\textrm{FO}}$ denotes the expansion of the
corresponding resummed endpoint contribution through the same
perturbative order in $\alpha_s$ as the fixed-order SDC in the
first term. The subtraction therefore removes precisely the
overlap between the resummed and fixed-order contributions,
avoiding double counting while retaining all fixed-order terms
not generated by the LL resummation.

\subsection{Resummation of threshold logarithms in soft-gluon factorization}

In SGF, the input FFs of Eq.~\eqref{eq:pqcdfac} are matched, at leading order in the velocity expansion and in the diagonal channel approximation used here, onto convolutions of perturbative matching coefficients with the soft-gluon distribution~(SGD) $F_{[n]\to H}$~\cite{Ma:2017xno,Chen:2021hzo}:
\begin{subequations}\label{eq:SGF-form}
\begin{align}
D_{f\to H}(z,\mu_0)
={}&\sum_n\int_z^1\frac{\md x}{x}
\hat D_{f\to n}(z/x)F_{[n]\to H}(x),\\
{\cal D}_{\kappa\to H}(z,\zeta,\zeta',\mu_0)
={}&\sum_n\int_z^1\frac{\md x}{x}
\hat{\cal D}_{\kappa\to n}(z/x,\zeta,\zeta')
 F_{[n]\to H}(x).
\end{align}
\end{subequations}
Here $x=p^+/P_c^+$ is the fraction of the intermediate-pair plus momentum $P_c^+$ carried by $H$, so the short-distance fragmentation variable is $z/x$. To keep Eq.~\eqref{eq:SGF-form} compact, the common arguments $M_H/x$, $\mu_0$, and $\mu_f$ of the matching coefficients and the arguments $M_H$, $m_Q$, and $\mu_f$ of the SGDs are suppressed. The quantities $\hat D$ and $\hat{\cal D}$ are perturbative SGF matching coefficients, $M_H$ is the quarkonium mass with $p^2=M_H^2$, and $\mu_f$ is the SGF factorization scale. The SGD describes the transition $Q\bar Q[n]\to H+X$; the LL soft-enhanced terms resummed in this work are assigned to it and summed through its renormalization-group evolution~\cite{Chen:2023gsu,Jia:2024cvv}.

The SGD is defined, schematically, by the operator matrix element
\begin{align}\label{eq:SGD-1d}
F_{[n] \to H}(x,M_H,m_Q,\mu_f)
&= p^+\int \frac{\mathrm{d}b^-}{2\pi} e^{-ip^+  b^-/x}  \langle 0| [\bar\Psi \mathcal {K}_{n} \Psi]^\dagger(0) [a_H^\dagger a_H] [\bar\Psi \mathcal {K}_{n}\Psi](b^-) |0\rangle_{\textrm{S}},
\end{align}
where $x=p^+/P_c^+$ is the light-cone momentum fraction, with $P_c^+$ the plus component of the total momentum of the intermediate $Q\bar Q$ pair, and $\Psi$ denotes the heavy-quark Dirac field. The subscript ``S'' indicates that the field operators are restricted to the small-momentum region; it also selects only the leading-power terms in the expansion in $(P_c-p)^+=(1-x)P_c^+$~\cite{Chen:2021hzo}. The operators $\mathcal {K}_{n}$ are projection
operators corresponding to the intermediate state $n$; their explicit definitions are given in Ref.~\cite{Ma:2017xno}. To lowest order in the velocity expansion, the normalization of the SGD is related to the LDME defined in Ref.~\cite{Ma:2015yka} as
\begin{align}\label{eq:SGD-LDME}
\int_0^1 \frac{dx}{x^2} F_{[n] \to H}(x,M_H,m_Q,\mu_f)
&\approx \langle \mathcal {O}^{H}(n)\rangle.
\end{align}

The SGDs satisfy, at leading order in the velocity expansion, the evolution equation~\cite{Chen:2021hzo}
\begin{align}\label{eq:general-RGE-SGDs}
\frac{d}{d \ln \mu_f}  F_{[n] \to H}(x,M_H,m_Q, \mu_f)
=& \sum_{n^\prime} \int_x^1 \frac{dy}{y} \boldsymbol{K}_{[n]}^{[n^\prime]}(x/y, M_H/y, \mu_f)
 F_{[n^\prime] \to H}(y,M_H,m_Q, \mu_f).
\end{align}
For $n=\COcSa$, $\COaSz$, and $\COcPj$, the LO kernel $\boldsymbol{K}_{[n]}^{[n^\prime]}$ has been derived in Refs.~\cite{Chen:2021hzo,Chen:2022qli} and reads
\begin{subequations}
\begin{align}
\boldsymbol{K}_{[\state{{3}}{S}{1,\lambda}{8}]}^{[\state{{3}}{S}{1,\lambda^\prime}{8}],\textrm{LO}}(x, M_H,\mu_f)=&\frac{\alpha_s }{4\pi} \biggl[ 2\Gamma_0^F \biggl( \frac{ x}{(1- x)_+}  -\ln \frac{ \mu_f }{M_H}
\delta(1- x)\biggr) + \gamma^F_0 \delta(1- x) \biggr] \delta_{\lambda \lambda^\prime}, \\
\boldsymbol{K}_{[\COaSz]}^{[\COaSz],\textrm{LO}}(x, M_H,\mu_f)=&\frac{\alpha_s }{4\pi} \biggl[ 2\Gamma_0^F \biggl( \frac{ x}{(1- x)_+}  -\ln \frac{ \mu_f }{M_H}
\delta(1- x)\biggr) +  \gamma^F_0 \delta(1- x) \biggr], \\
\boldsymbol{K}_{[\state{{3}}{P}{J,\lambda}{8}]}^{[\state{{3}}{P}{J^\prime,\lambda^\prime}{8}],\textrm{LO}}(x, M_H,\mu_f)=&\frac{\alpha_s }{4\pi} \biggl[ 2\Gamma_0^F \biggl( \frac{ x}{(1- x)_+}  -\ln \frac{ \mu_f }{M_H}
\delta(1- x)\biggr) + \gamma^F_0 \delta(1- x) \biggr] \delta_{JJ^\prime}\delta_{\lambda \lambda^\prime},
\end{align}
\end{subequations}
where
\begin{align}\label{eq:cusp}
\Gamma^F_0 = 4 N_c,  \quad \gamma^F_0 =& 4 N_c .
\end{align}
To solve the corresponding DGLAP equation for the SGDs, we work directly in Mellin space. For the $\COcSa$ SGD, the transformed equation becomes
\begin{align}\label{eq:MellinRGE}
\frac{d}{d \ln\mu_f}
\tilde F_{[\COcSa] \to H}(N, M_H,m_Q, \mu_f)
=&  \frac{\alpha_s}{4\pi} \biggl(  -2\Gamma^F_{0}   \ln \frac{\bar N \mu_f}{M_H}
+ \gamma^F_0 \biggr) \tilde F_{[\COcSa] \to H}(N, M_H,m_Q, \mu_f)+ \mathcal{O}\Big(\frac{1}{N}\Big),
\end{align}
where $\bar N=N e^{\gamma_E}$ and $\gamma_E$ is Euler's constant. Here we retain only the threshold-logarithmic terms. Evolving the SGD from the characteristic scale $M_H/N$ to $M_H$, we obtain~\cite{Chen:2021hzo,Chen:2022qli}
\begin{align}\label{eq:Resummed-SGD}
\tilde {F}_{[\COcSa] \to H}(N,M_H,m_Q, M_H) =&
  \mathrm{exp}\Big[ h_0(\chi_{0}) \Big] \tilde {F}_{[\COcSa] \to H}(N,M_H,m_Q, M_H/N).
\end{align}
Here
\begin{align}\label{eq:sgfh}
h_0(\chi_{0})=&\frac{2\pi \Gamma^F_0}{\beta_0^2}   \frac{ 1}{\alpha_s(M_H)} \Big(-(1-2\chi_{0})\ln (1-2\chi_{0}) - 2\chi_{0} \Big)
- \frac{ \gamma_0^F}{2\beta_0} \ln (1-2\chi_{0}) + \frac{ \Gamma_0^F \gamma_E}{\beta_0} \ln (1-2\chi_{0}),
\end{align}
with
\begin{align}\label{eq:chi}
\chi_{0} = \frac{\alpha_s(M_H) \beta_0}{4\pi}\ln N, \quad\quad\quad   \beta_0=(11/3)C_A-(4/3)T_F n_f.
\end{align}
Here $n_f$ is the number of active quark flavors, and we set $n_f=3$. For SU(3), the color factors are $T_F=1/2$, $C_F=4/3$, and $C_A=3$. The logarithm $\ln(1-2\chi_0)$ in Eq.~\eqref{eq:sgfh} produces a Landau singularity at the branch point
\begin{align}
N^L = \exp\biggl(\frac{2\pi}{\beta_0 \alpha_s(M_H)} \biggr).
\end{align}
Following Refs.~\cite{Cacciari:2005uk,Shimizu:2005fp}, we avoid this singularity through the replacement
\begin{align}
\chi_0 \to \chi_{0\ast} = \frac{\alpha_s(M_H) \beta_0}{4\pi}\ln  \left(\frac{N}{1+N/N_{\ast}^L}\right), \quad N_{\ast}^L=N^L/a,
\end{align}
where $a$ is an order-one parameter with $a\ge 1$. This replacement keeps $\chi_0$ out of the nonperturbative region, whose effects are instead absorbed into the initial SGD. We parametrize that initial input as
\begin{align}\label{eq:model}
\tilde {F}_{[n] \to H}(N,M_H,m_Q, M_H/N) =&
  \sum_{n^\prime} \tilde{\boldsymbol{D}}_{[n]}^{[n^\prime]}(N,M_H,\mu_\Lambda) \tilde {F}_{[n^\prime] \to H}^{\textrm{mod}}(N,M_H).
\end{align}
The coefficient $\tilde{\boldsymbol{D}}_{[n]}^{[n^\prime]}$ is fixed by the matching relation
\begin{align}\label{eq:coefficient}
 \tilde {F}_{[n] \to Q\bar Q[m]}(N,M_H,m_Q, M_H/N)\Big\vert_{m_Q=M_H/2}
=& \sum_{n^\prime} \tilde{\boldsymbol{D}}_{[n]}^{[n^\prime]}(N,M_H,\mu_\Lambda)  \langle \mathcal {O}^{Q\bar Q[m]}(n^\prime)\rangle(\mu_\Lambda)\Big\vert_{m_Q=M_H/2}
.
\end{align}
The subscript $m_Q=M_H/2$ indicates that only the leading term in the velocity expansion is retained. The coefficient $\tilde{\boldsymbol{D}}_{[n]}^{[n^\prime]}$ accounts for perturbative effects in $\tilde{F}_{[n] \to Q\bar Q[n^\prime]}(N,M_H,m_Q, M_H/N)$ away from the strict endpoint. Combining Eq.~\eqref{eq:coefficient} with the perturbative SGD $F_{[n] \to Q\bar Q[m]}$ computed in Ref.~\cite{Chen:2023gsu}, we obtain the LO expressions in $x$ space relevant to $J/\psi$ production:
\begin{subequations}\label{eq:SGD-result}
\begin{align}
\boldsymbol{D}_{[\state{{3}}{S}{1,\lambda}{8}]}^{[\state{{3}}{S}{1,\lambda^\prime}{8}],\textrm{LO}}(x, M_H,\mu_\Lambda)=&\delta(1-x) \delta_{\lambda \lambda^\prime}, \\
\boldsymbol{D}_{[\COaSz]}^{[\COaSz],\textrm{LO}}(x, M_H,\mu_\Lambda)=&\delta(1-x), \\
\boldsymbol{D}_{[\state{{3}}{P}{J,\lambda}{1}]}^{[\state{{3}}{P}{J^\prime,\lambda^\prime}{1}],\textrm{LO}}(x, M_H,\mu_\Lambda)=&\delta(1-x) \delta_{JJ^\prime}\delta_{\lambda \lambda^\prime},\\
\boldsymbol{D}_{[\state{{3}}{P}{J,\lambda}{8}]}^{[\state{{3}}{P}{J^\prime,\lambda^\prime}{8}],\textrm{LO}}(x, M_H,\mu_\Lambda)=&\delta(1-x) \delta_{JJ^\prime}\delta_{\lambda \lambda^\prime},\\
\boldsymbol{D}_{[\state{{3}}{S}{1,T}{8}]}^{[\state{{3}}{P}{0}{1}],\textrm{LO}}(x,M_H,\mu_\Lambda)
=&  \frac{\alpha_s(M_H) }{ M_H^2 \pi }\frac{N_c^2-1}{N_c}\frac{8}{9}
        \Big[ \Big( - \ln \frac{  \mu_\Lambda^2 }{M_H^2} + \frac{1}{2} \Big) \delta(1-x)
        +  \frac{2x}{(1-x)_+} \Big], \\
\boldsymbol{D}_{[\state{{3}}{S}{1,T}{8}]}^{[\state{{3}}{P}{1,T}{1}],\textrm{LO}}(x,M_H,\mu_\Lambda)
=&  \frac{\alpha_s(M_H) }{ M_H^2 \pi }\frac{N_c^2-1}{N_c}\frac{4}{3}
        \Big[  -  \ln \frac{  \mu_\Lambda^2 }{M_H^2}
        \delta(1-x)
        +  \frac{2x}{(1-x)_+} \Big],\\
\boldsymbol{D}_{[\state{{3}}{S}{1,T}{8}]}^ {[\state{{3}}{P}{1,L}{1}],\textrm{LO}}(x,M_H,\mu_\Lambda)
=&  \frac{\alpha_s(M_H) }{ M_H^2 \pi }\frac{N_c^2-1}{N_c}\frac{4}{3}
        \Big[ \Big( -  \ln \frac{  \mu_\Lambda^2 }{M_H^2} + \frac{1}{2} \Big)
        \delta(1-x)+  \frac{2x}{(1-x)_+} \Big] ,\\
\boldsymbol{D}_{[\state{{3}}{S}{1,T}{8}]}^{[\state{{3}}{P}{2,TT}{1}],\textrm{LO}}(x,M_H,\mu_\Lambda)
=&  \frac{\alpha_s(M_H) }{ M_H^2 \pi }\frac{N_c^2-1}{N_c}\frac{8}{3}
        \Big[ \Big( -  \ln \frac{  \mu_\Lambda^2 }{M_H^2} + \frac{1}{2} \Big)
        \delta(1-x)
        +  \frac{2x}{(1-x)_+} \Big]  ,\\
\boldsymbol{D}_{[\state{{3}}{S}{1,T}{8}]}^{[\state{{3}}{P}{2,T}{1}],\textrm{LO}}(x,M_H,\mu_\Lambda)
=&  \frac{\alpha_s(M_H) }{ M_H^2 \pi }\frac{N_c^2-1}{N_c}\frac{4}{3}
        \Big[  -  \ln \frac{  \mu_\Lambda^2 }{M_H^2}
        \delta(1-x)
        +  \frac{2x}{(1-x)_+} \Big] ,\\
\boldsymbol{D}_{[\state{{3}}{S}{1,T}{8}]} ^{[\state{{3}}{P}{2,L}{1}],\textrm{LO}}(x,M_H,\mu_\Lambda)
=&  \frac{\alpha_s(M_H) }{ M_H^2 \pi }\frac{N_c^2-1}{N_c}\frac{4}{9}
        \Big[ \Big( -  \ln \frac{  \mu_\Lambda^2 }{M_H^2} + \frac{1}{2} \Big)
        \delta(1-x)
        +  \frac{2x}{(1-x)_+} \Big],\\
\boldsymbol{D}_{[\state{{3}}{S}{1,T}{8}]} ^{[\state{{3,T}}{P}{}{1}],\textrm{LO}}(x,M_H,\mu_\Lambda)
=&  \frac{\alpha_s(M_H) }{ M_H^2 \pi }\frac{N_c^2-1}{N_c}8
        \Big[ \Big( -  \ln \frac{  \mu_\Lambda^2 }{M_H^2} + \frac{1}{3} \Big)
        \delta(1-x)
        +  \frac{2x}{(1-x)_+} \Big],\\
\boldsymbol{D}_{[\state{{3}}{S}{1,T}{8}]} ^{[\state{{3,L}}{P}{}{1}],\textrm{LO}}(x,M_H,\mu_\Lambda)
=&  0,\\
\boldsymbol{D}_{[\state{{3}}{S}{1,T}{8}]} ^{[\state{{3}}{P}{J,\lambda}{8}],\textrm{LO}}(x,M_H,\mu_\Lambda)
=& \frac{N_c^2-4}{2(N_c^2-1)} \boldsymbol{D}_{[\state{{3}}{S}{1,T}{8}]} ^{[\state{{3}}{P}{J,\lambda}{1}],\textrm{LO}}(x,M_H,\mu_\Lambda).
\end{align}
\end{subequations}
Here $\state{{3,\lambda}}{P}{}{1,8}$ denotes the $\state{{3}}{P}{J}{1,8}$ channel summed over the total angular momentum $J$ at fixed helicity~\cite{Ma:2015yka}: the labels $\lambda=L, T, TT, \cdots$ correspond to $|J_z|=0,1,2,\cdots$, respectively.

For the nonperturbative model \(\tilde {F}_{[n^\prime] \to H}^{\textrm{mod}}\), we adopt the parametrization introduced in Ref.~\cite{Chen:2022qli}. In momentum space, it reads
\begin{align}\label{eq:model-SGD-SW}
F_{[n^\prime] \to H}^{\textrm{mod}}(x,M_H) = \langle \mathcal {O}^{H}(n^\prime)\rangle \frac{ M_H}{\Gamma(b)}\frac{(M_H \omega)^{b-1}}{(\bar{\Lambda}[n^\prime])^b} \exp\Big(-\frac{M_H \omega}{\bar{\Lambda}[n^\prime]}\Big), \quad \omega = \frac{1}{x} - 1,
\end{align}
where the normalization, first moment, and second moment are given by
\begin{subequations}\label{eq:moments}
\begin{align}
& \int_0^1 \frac{dx}{x^2} F_{[n^\prime] \to H}^{\textrm{mod}}(x,M_H)=\langle \mathcal {O}^{H}(n^\prime)\rangle,\\
& \int_0^1 \frac{dx}{x^2} M_H\Big(\frac{1}{x}-1\Big) F_{[n^\prime] \to H}^{\textrm{mod}}(x,M_H)= b\bar{\Lambda}[n^\prime] \langle \mathcal {O}^{H}(n^\prime)\rangle,\\
& \int_0^1 \frac{dx}{x^2} M_H^2\Big(\frac{1}{x}-1\Big)^2 F_{[n^\prime] \to H}^{\textrm{mod}}(x,M_H)= b(b+1) (\bar{\Lambda}[n^\prime])^2\langle \mathcal {O}^{H}(n^\prime)\rangle.
\end{align}
\end{subequations}
The parameter $b \bar{\Lambda}[n^\prime]$ characterizes the average momentum radiated in the hadronization process $Q\bar Q[n^\prime]\to H$. Following Refs.~\cite{Fleming:2003gt,Chen:2022qli}, we choose $a=1.1$ and $b=2$. For $H=J/\psi$ and $\psi(2S)$, we set $2\bar{\Lambda}[\COaSz]=2\bar{\Lambda}[\COcPz]=0.6~\textrm{GeV}$ and $2\bar{\Lambda}[\COcSa]=1.2~\textrm{GeV}$. For $H=\chi_{cJ}$ and $\eta_c$, we use $2\bar{\Lambda}[\CScPz]=0.3~\textrm{GeV}$ and $2\bar{\Lambda}[\COcSa]=0.6~\textrm{GeV}$.

Solving the RGE for $F_{[\COcSa] \to H}$ with Eqs.~\eqref{eq:general-RGE-SGDs},~\eqref{eq:MellinRGE},~\eqref{eq:model}, and~\eqref{eq:SGD-result}, we find that the leading threshold logarithms for the $\state{{3}}{P}{J,\lambda}{1,8}$ and $\COcSa$ gluon FFs are resummed by the same evolution function, $h_0(\chi_0)$. Our LL expression for the $\COcSa$ gluon FF agrees with that of Ref.~\cite{Chung:2024jfk}, but the LL terms for the $\state{{3}}{P}{J}{1,8}$ channels do not. The difference originates in the structure of the evolution equation: in Eq.~\eqref{eq:general-RGE-SGDs} we retain only the $\mathcal O(\alpha_s)$ kernels $\boldsymbol{K}_{[n]}^{[n^\prime]}$, omitting the $\mathcal O(\alpha_s^2)$ mixing kernels such as $\boldsymbol{K}_{[\state{{3}}{S}{1,\lambda}{8}]}^{[\state{{3}}{P}{J,\lambda^\prime}{1,8}]}$, whereas Ref.~\cite{Chung:2024jfk} treats $\mathcal O(\alpha_s)$ and $\mathcal O(\alpha_s^2)$ contributions on the same footing within the evolution functions.

\subsection[Fit results for chi cJ, psi(2S), and eta c production]{Fit results for $\chi_{cJ}$, $\psi(2S)$, and $\eta_c$ production}
We collect here the auxiliary fits for $\chi_{cJ}$, $\psi(2S)$, and $\eta_c$ hadroproduction that support the interpretation in the main text and show how the same mechanism extends beyond $J/\psi$.

For $\chi_{cJ}$ production, we fit the ATLAS measurements of the $\chi_{c1}$ and $\chi_{c2}$ transverse-momentum spectra at $\sqrt{s}=7~\rm TeV$~\cite{ATLAS:2014ala}. The $\chi_{c1}$ and $\chi_{c2}$ matrix elements are related to those of $\chi_{c0}$ by heavy-quark spin symmetry, $\langle \mathcal{O}^{\chi_{cJ}}(n)\rangle=(2J+1)\langle \mathcal{O}^{\chi_{c0}}(n)\rangle[1+\mathcal{O}(v^2)]$, so the fit determines the two LDMEs $\langle \mathcal{O}^{\chi_{c0}}(\COcSa)\rangle$ and $\langle \mathcal{O}^{\chi_{c0}}(\CScPz)\rangle$. The fit gives
\begin{subequations}\label{eq:NREM}
\begin{align}
\langle \mathcal{O}^{\chi_{c0}}(\COcSa)\rangle\vert_{\textrm{NRQCD}}=&(1.186\pm 0.219)\times 10^{-2} \textrm{GeV}^3,\\
\frac{\langle \mathcal{O}^{\chi_{c0}}(\CScPz)\rangle}{m_c^2}\vert_{\textrm{NRQCD}}=&(5.683\pm 1.900)\times 10^{-3} \textrm{GeV}^3,
\end{align}
\end{subequations}
with $\chi^2=0.5$ for 8 degrees of freedom in NRQCD factorization, and
\begin{subequations}\label{eq:SGFEM}
\begin{align}
\langle \mathcal{O}^{\chi_{c0}}(\COcSa)\rangle\vert_{\textrm{SGF}}=&(2.568\pm 0.425)\times 10^{-2} \textrm{GeV}^3,\\
\frac{\langle \mathcal{O}^{\chi_{c0}}(\CScPz)\rangle}{m_c^2}\vert_{\textrm{SGF}}=&(2.287\pm 0.765)\times 10^{-2} \textrm{GeV}^3,
\end{align}
\end{subequations}
with $\chi^2=0.57$ for 8 degrees of freedom in SGF. Figure~\ref{fig:NRatlas} shows the corresponding $\chi_{c1}$ and $\chi_{c2}$ spectra, decomposed into the separate channel contributions, together with the ATLAS data. The resummed framework thus also provides a stable description of the $\chi_{cJ}$ system.
\begin{figure*}
\begin{center}
\begin{tabular}{cc}
\includegraphics[width=0.45\textwidth]{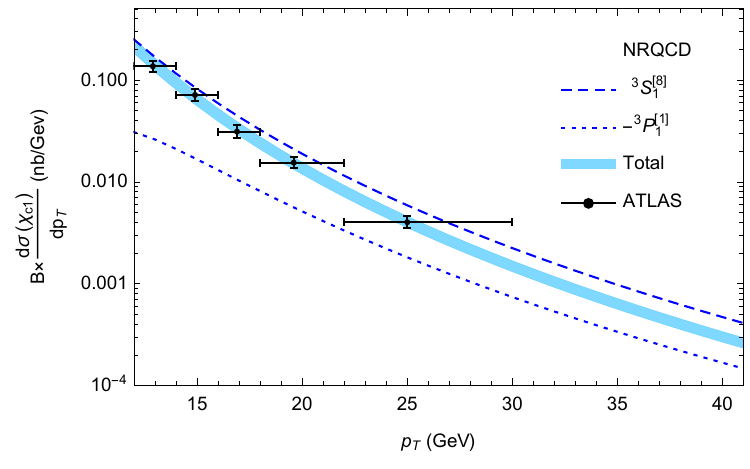}
&
\includegraphics[width=0.45\textwidth]{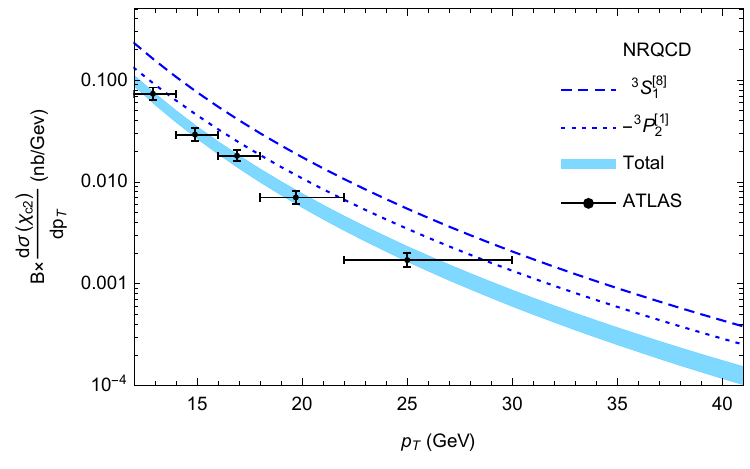}
\\
\includegraphics[width=0.45\textwidth]{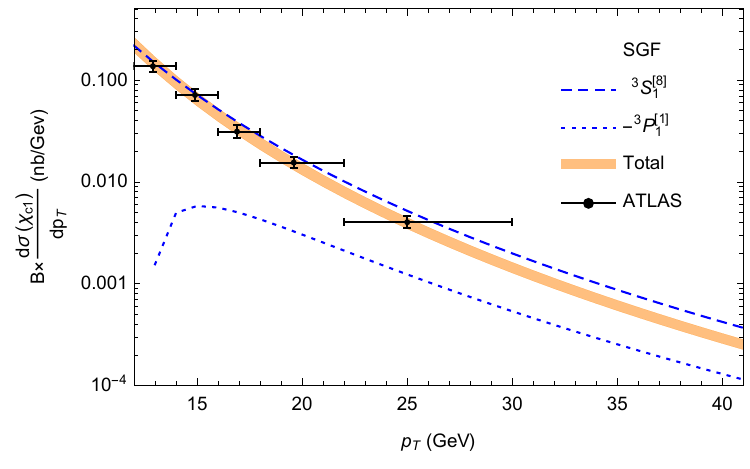}
&
\includegraphics[width=0.45\textwidth]{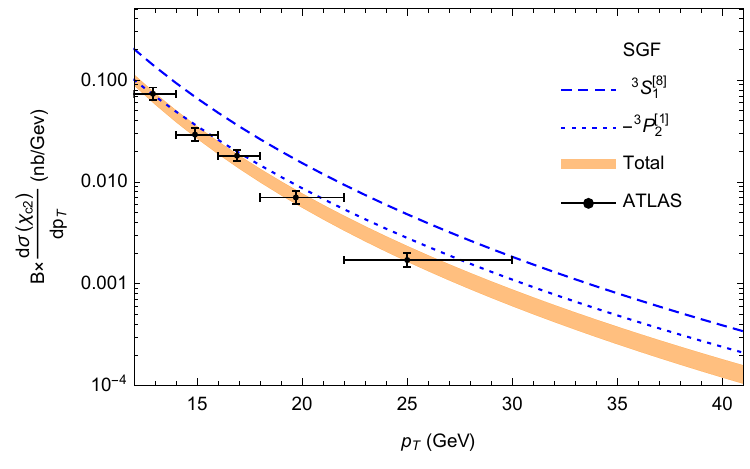}
\end{tabular}
\end{center}
\caption{Upper (lower) row: results in NRQCD (SGF). Left (right) column: prompt $\chi_{c1}$ ($\chi_{c2}$) differential cross sections at $\sqrt{s}=7~\rm TeV$ and $|y|<0.75$, compared with ATLAS measurements~\cite{ATLAS:2014ala}. The factor $\textrm{B}$ denotes the product of branching fractions, $\textrm{B}=\textrm{Br}(\chi_{cJ}\to J/\psi + \gamma)\textrm{Br}(J/\psi \to \mu^+\mu^-)$.}\label{fig:NRatlas}
\end{figure*}

For $\psi(2S)$, we fit the unpolarized yield data from ATLAS and CMS~\cite{ATLAS:2014zpz,CMS:2015lbl}. As in the $J/\psi$ analysis, only two linear combinations of the three CO LDMEs are well constrained:
\begin{subequations}\label{eq:psi2S-combination}
\begin{align}
&M_{r_0^\prime}^{\psi(2S)}\vert_{X} \equiv \langle \mathcal{O}^{\psi(2S)}(\COaSz)\rangle + r_0^{\prime X} \frac{\langle \mathcal{O}^{\psi(2S)}(\COcPz)\rangle}{m_c^2}, \\
&M_{r_1^\prime}^{\psi(2S)}\vert_{X} \equiv \langle \mathcal{O}^{\psi(2S)}(\COcSa)\rangle + r_1^{\prime X} \frac{\langle \mathcal{O}^{\psi(2S)}(\COcPz)\rangle}{m_c^2},
\end{align}
\end{subequations}
where $X\in\{\textrm{NRQCD},\textrm{SGF}\}$, with projection coefficients $r_0^{\prime \textrm{NRQCD}} = 22.45$ and $r_1^{\prime \textrm{NRQCD}} = -5.16$ in NRQCD, and $r_0^{\prime\textrm{SGF}} = 3.47$ and $r_1^{\prime\textrm{SGF}} = -2.09$ in SGF. The fit yields
\begin{subequations}\label{eq:psi2S-M0M1}
\begin{align}
M_{r_0^\prime}^{\psi(2S)}\vert_{\textrm{NRQCD}} =& (1.663\pm 0.418)\times 10^{-1} \textrm{GeV}^3, \\
M_{r_1^\prime}^{\psi(2S)}\vert_{\textrm{NRQCD}} =&(-3.074\pm 3.106)\times 10^{-3} \textrm{GeV}^3,
\end{align}
\end{subequations}
in NRQCD factorization with $\chi^2=16.0$ for 86 degrees of freedom, and
\begin{subequations}\label{eq:psi2S-SGF-M0M1}
\begin{align}
M_{r_0^\prime}^{\psi(2S)}\vert_{\textrm{SGF}} =& (1.783\pm 0.404)\times 10^{-1} \textrm{GeV}^3, \\
M_{r_1^\prime}^{\psi(2S)}\vert_{\textrm{SGF}} =&(3.130\pm 0.380)\times 10^{-2} \textrm{GeV}^3,
\end{align}
\end{subequations}
in SGF with $\chi^2=14.5$ for 86 degrees of freedom. Fig.~\ref{fig:atlas-cms} compares the fitted spectra with the data.
\begin{figure*}
\begin{center}
\begin{tabular}{cc}
\includegraphics[width=0.45\textwidth]{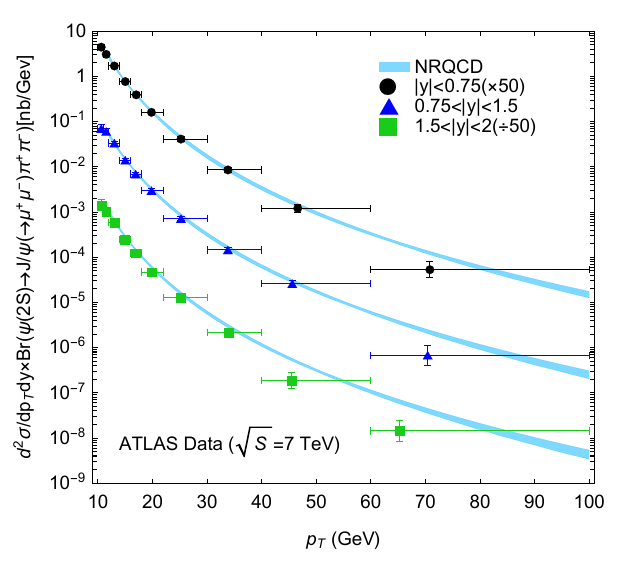}
&
\includegraphics[width=0.45\textwidth]{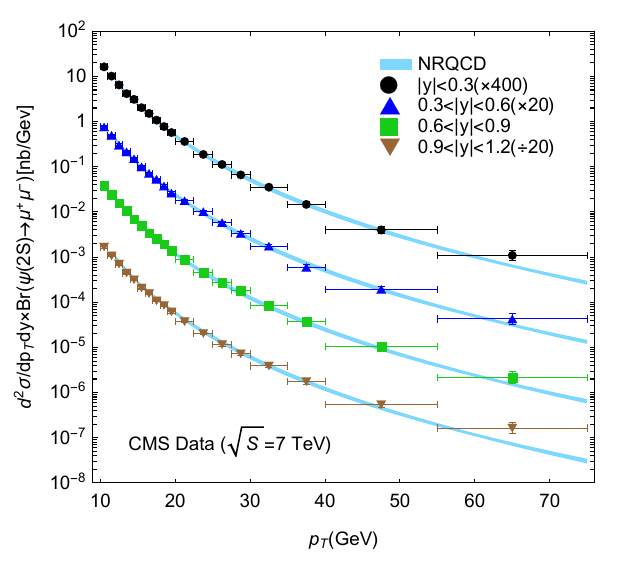}
\\
\includegraphics[width=0.45\textwidth]{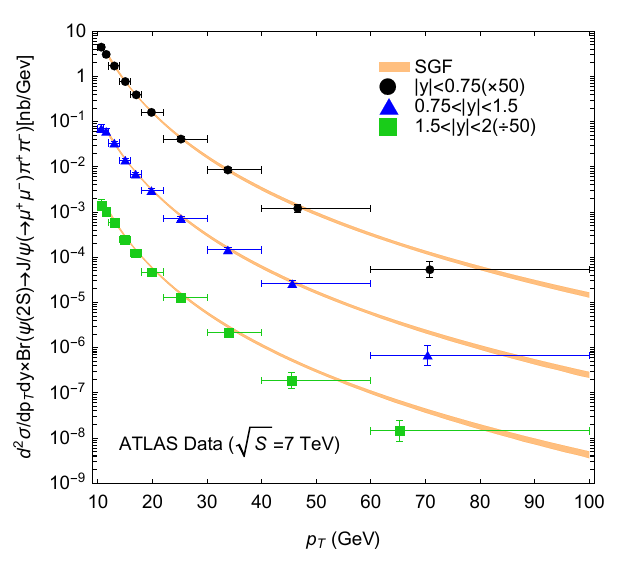}
&
\includegraphics[width=0.45\textwidth]{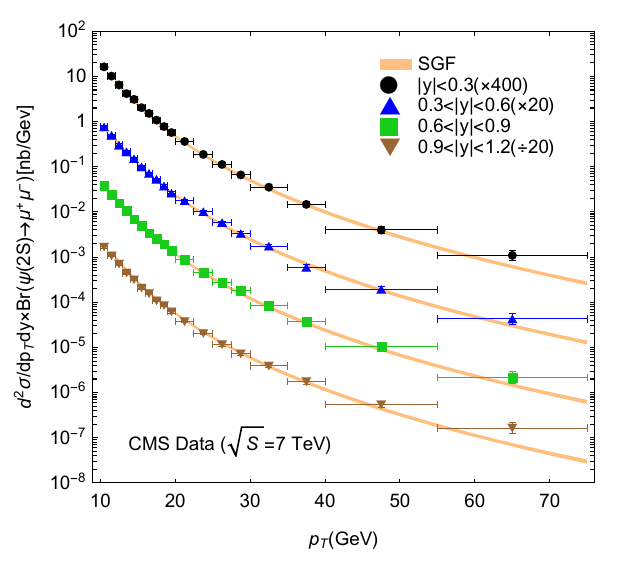}
\end{tabular}
\end{center}
\caption{Upper (lower) row: results in NRQCD (SGF). Left (right) column: fitted $\psi(2S)$ yield compared with ATLAS (CMS) data~\cite{ATLAS:2014zpz,CMS:2015lbl}.}\label{fig:atlas-cms}
\end{figure*}
The description deteriorates at the largest $p_T$, where the ATLAS and CMS measurements themselves differ. Figure~\ref{fig:psi2S-lambda} shows the $\psi(2S)$ polarization parameter $\lambda_\theta$ for four representative values of $\xi$, in both NRQCD and SGF. As in the $J/\psi$ case, the polarization is highly sensitive to the spin-flip contribution, and with $\xi$ of order $v^4$ the predictions agree with the CMS measurements~\cite{CMS:2013gbz}. The same cancellation mechanism is therefore operative beyond the ground state.
\begin{figure*}
\begin{center}
\begin{tabular}{cccc}
\includegraphics[width=0.25\textwidth]{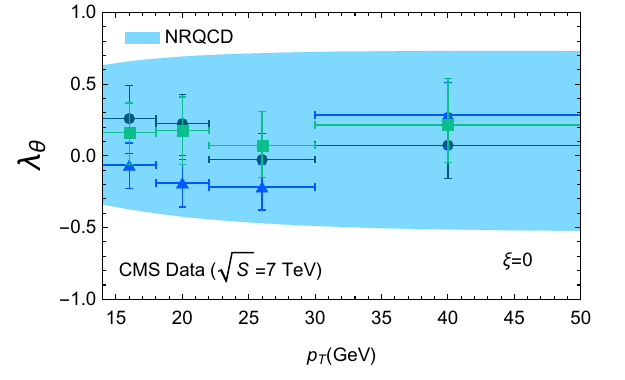}
&
\includegraphics[width=0.25\textwidth]{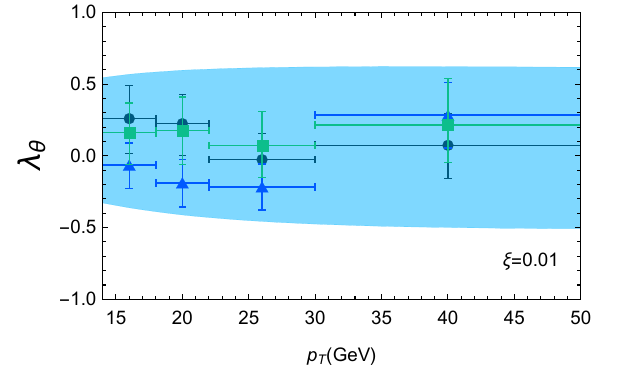}
&
\includegraphics[width=0.25\textwidth]{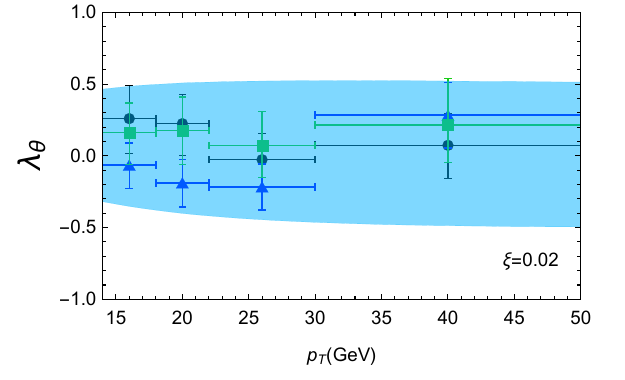}
&
\includegraphics[width=0.25\textwidth]{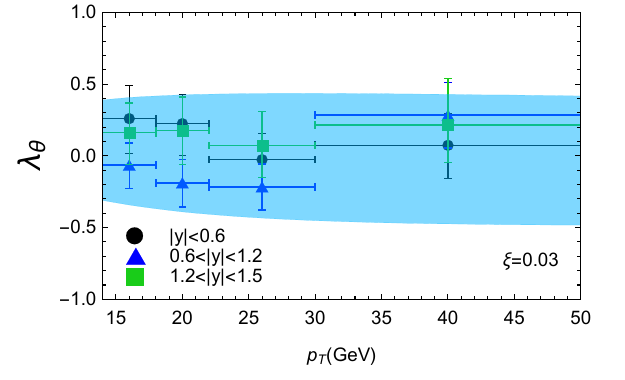}
\\
\includegraphics[width=0.25\textwidth]{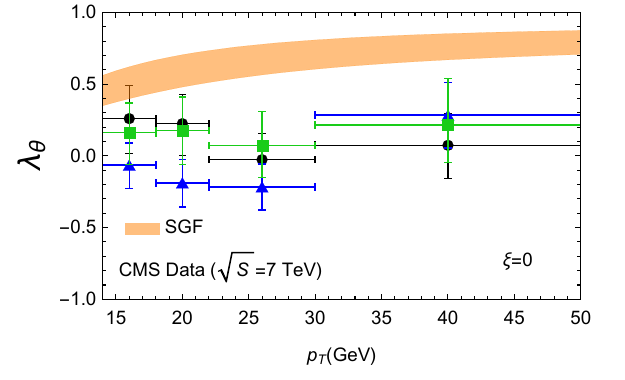}
&
\includegraphics[width=0.25\textwidth]{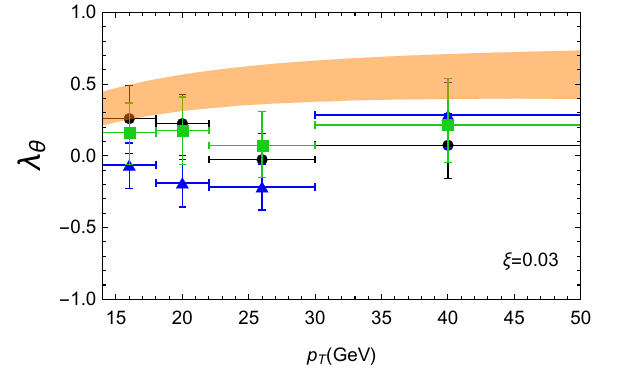}
&
\includegraphics[width=0.25\textwidth]{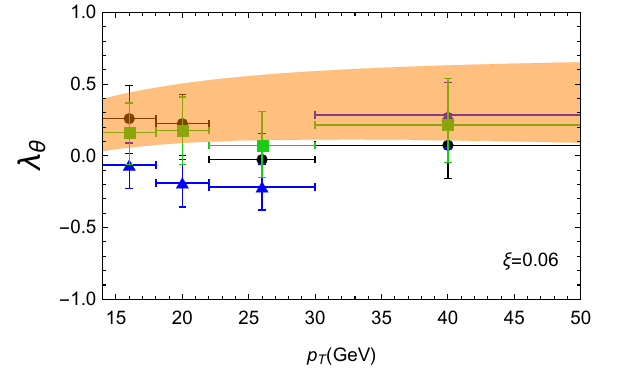}
&
\includegraphics[width=0.25\textwidth]{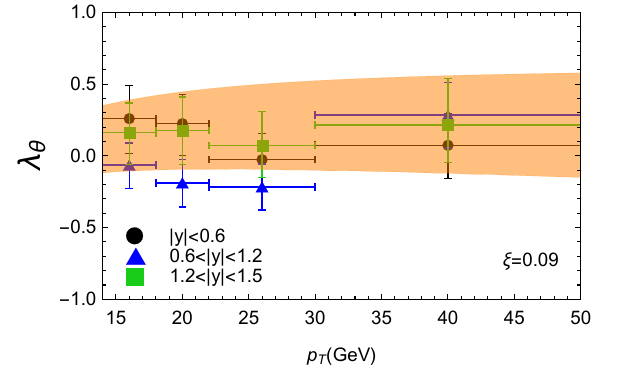}
\end{tabular}
\end{center}
\caption{Upper (lower) row: results in NRQCD (SGF). The four columns correspond, from left to right, to four representative values of $\xi$. The curves show the predicted $\psi(2S)$ polarization parameter $\lambda_\theta$ in the helicity frame for $|y|<0.6$, compared with CMS data~\cite{CMS:2013gbz}.}\label{fig:psi2S-lambda}
\end{figure*}

For $\eta_c$ production, we follow Ref.~\cite{Han:2014jya} and use the LHCb data~\cite{LHCb:2014oii} to constrain $\langle \mathcal{O}^{J/\psi}(\COaSz)\rangle$ through heavy-quark spin symmetry. In the LHCb kinematic window, the dominant contributions come from the $\CSaSz$ and $\COcSa$ channels~\cite{Han:2014jya,Zhang:2014ybe}. Saturating the data with the $\COcSa$ channel alone, following Ref.~\cite{Han:2014jya}, we find
\begin{subequations}\label{eq:eta-EM}
\begin{align}
\langle\mathcal{O}^{\eta_c}(\COcSa)\rangle\vert_{\textrm{NRQCD}} =& (6.854 \pm 2.611) \times 10^{-2} \ \textrm{GeV}^3,\\
\langle\mathcal{O}^{\eta_c}(\COcSa)\rangle\vert_{\textrm{SGF}} =& (1.236 \pm 0.471) \times 10^{-1} \ \textrm{GeV}^3.
\end{align}
\end{subequations}
Interpreting saturation as an upper limit, we take the fitted central value as the bound and set the lower limit to zero by positivity of the LDME~\cite{Han:2014jya,Shao:2014yta}:
\begin{subequations}\label{eq:EM-bound}
\begin{align}
0 < \langle\mathcal{O}^{\eta_c}(\COcSa)\rangle\vert_{\textrm{NRQCD}} <& 6.854 \times 10^{-2} \ \textrm{GeV}^3,\\
0 < \langle\mathcal{O}^{\eta_c}(\COcSa)\rangle\vert_{\textrm{SGF}} <& 1.236 \times 10^{-1} \ \textrm{GeV}^3.
\end{align}
\end{subequations}
Using the heavy-quark spin-symmetry relation~\cite{Bodwin:1994jh}
\begin{align}\label{eq:HQSS}
 \langle\mathcal{O}^{\eta_c}(\COcSa)\rangle \approx \langle\mathcal{O}^{J/\psi}(\COaSz)\rangle,
\end{align}
we obtain the corresponding constraint on $\langle\mathcal{O}^{J/\psi}(\COaSz)\rangle$:
\begin{subequations}\label{eq:eta-constraint}
\begin{align}
0 < \langle\mathcal{O}^{J/\psi}(\COaSz)\rangle\vert_{\textrm{NRQCD}} <& 6.854 \times 10^{-2} \ \textrm{GeV}^3,\\
0 < \langle\mathcal{O}^{J/\psi}(\COaSz)\rangle\vert_{\textrm{SGF}} <& 1.236 \times 10^{-1} \ \textrm{GeV}^3.
\end{align}
\end{subequations}
These bounds are weaker than, and therefore consistent with, the independent $J/\psi$ constraints in Eqs.~\eqref{eq:NRQCD-constraint} and~\eqref{eq:SGF-constraint}; they provide a complementary check of the allowed region.

\end{widetext}
\end{document}